# Debye temperature, electron-phonon coupling constant, and three-dome shape of crystalline strain as a function of pressure in highly compressed $La_3Ni_2O_{7-\delta}$

Talantsev Evgeny F.[†,1], Chistyakov Vasiliy V.[1]

[†]e-mail evgeny.talantsev@imp.uran.ru

[1]M.N. Miheev Institute of Metal Physics, Ural Branch, Russian Academy of Sciences, 18, S. Kovalevskoy St., Ekaterinburg, 620108, Russia

**Abstract.** Besides ongoing studies of phase structural transitions, pairing mechanism, and physical properties of recently discovered highly compressed high-temperature superconductor $La_3Ni_2O_{7-\delta}$, here we explored a possibility for the electron-phonon pairing mechanism as an origin of the superconducting state and determined the microcrystalline strain, ε, in high-pressure *Fmmm*-phase, and low-pressure *Amam*-phase of this nickelate. To do this, we analysed temperature dependent resistance and extracted pressure dependent Debye temperature, $\Theta_D(P)$, in $La_3Ni_2O_{7-\delta}$ with an approximate value of $\Theta_D(25\ GPa) = 550\ K$. From this we established that the $La_3Ni_2O_{7-\delta}$ is strong-coupled superconductor with the electron-phonon coupling constant $\lambda_{e-ph}(P=22.4\ GPa) = 1.75$. This value is close to $\lambda_{e-ph} = 1.70$ of ambient pressure superconductors $Nb_3R$ (R = Sn, Al). To address ongoing discussion that the lattice strain can be the origin for the emergence of high-temperature superconductivity in the $La_3Ni_2O_{7-\delta}$, we determined the microcrystalline strain, $0.011 \leq \varepsilon(P)$, in the high-pressure *Fmmm*-phase, and $\varepsilon(P) < 0.011$ of low-pressure *Amam*-phase. Our analysis showed that $\varepsilon(P)$ has three-dome shape in the pressure range of $1.6\ GPa \leq P \leq 41.2\ GPa$. One of these two $\varepsilon(P)$ deeps at $P = 15\ GPa$ coincides with the pressure at which the *Amam* into the *Fmmm* phase transition occurs. Based on our analysis, we proposed probable condition to observe the zero-resistance state in $La_3Ni_2O_{7-\delta}$.

## 1. Introduction

Experimental discovery of superconductivity with a transition temperature above 200 K in highly compressed sulphur hydride by Drozdov *et al* [1] manifested a new era in superconductivity. In the following years from this pivotal discovery[1], dozens of superconducting hydride phases have been discovered [2–4] and several fundamental effects have been recorded [5–7]. Fascinating feature of this research field is that experimental and first-principles calculations (FPC) quests in exploring the ultimate upper limit in superconductivity are in close collaboration [8].

High-temperature superconductivity in nickelates was predicted by Anisimov *et al*[9], and experimental discovery of the superconductivity with transition temperature $T_{c,zero} \sim 15\ K$ in $Nd_{0.80}Sr_{0.20}NiO_2$ thin films were reported by Li *et al*[10]. Recently, the family of highly pressurized superconductors was extended by another nickelate phase [11–14], $La_3Ni_2O_7$, which surpasses other nickelate phases with the highest $T_c$. Phase structural transition, pairing mechanism, superconducting gap symmetry and other properties/parameters of the $La_3Ni_2O_7$ are under ongoing theoretical and experimental investigations.

We need to mention that that there are some current experimental and theoretical challenges to explain the superconductivity in the nickelates. For instance, doped $RNiO_2$ (R = La, Pr, Nd) exhibits superconductivity with $T_{c,zero} \geq 1.9\ K$, but only in thin films of several nanometers thick. Bulk samples do not exhibit any sign of superconductivity down to $T_{c,onset} < 1.9\ K$ [15]. However, researchers very rarely mention this problem in the majority of theoretical and experimental studies, leading to a lack of understanding regarding the primary mechanism for the superconductivity in the nickelates.

Another example is the superconductivity in atomically thin quintuple-layer square-planar nickelate superlattice[16], for which only a single $\rho(T)$ dataset has been revealed. In Fig. 1 we showed a low-temperature part of this $\rho(T)$ dataset.

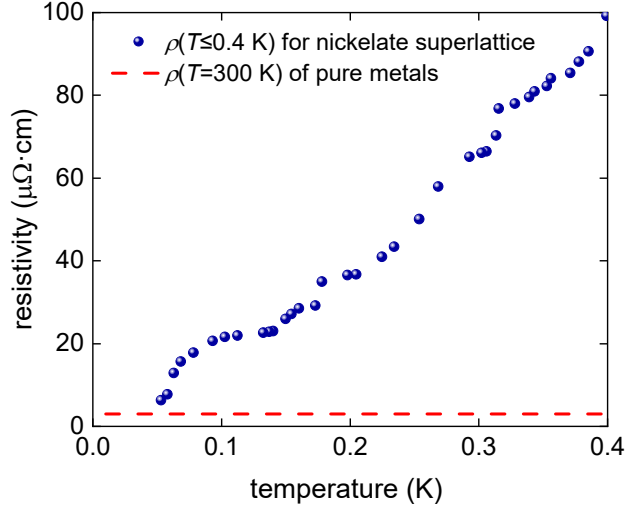

Fig. 1. Temperature dependent resistivity, $\rho(T \leq 0.4\ K)$, reported in quintuple-layer square-planar nickelate $Nd_6Ni_5O_{12}$ (raw data reported by Pan *et al*[16]). Red dash line of $\rho(T = 300\ K) = 3\ \mu\Omega cm$ is typical value for pure metals (see, for instance, Ref.[17]).

Simple examination of this experimental data shows that the $Nd_6Ni_5O_{12}$ exhibits the lowest measured resistance:
$$\rho(T = 53\ mK) = 6.3\ \mu\Omega \times cm \qquad (1)$$
which is higher than the resistivity of practically all pure metals at room temperature[17]:
$$\rho(T = 300\ K) < 6.3\ \mu\Omega \times cm \qquad (2)$$
In addition, Pan *et al.*[16] did not report the uncertainty level of the measurements, thus, we cannot agree that these authors observed the superconducting state in quintuple-layer square-planar nickelate $Nd_6Ni_5O_{12}$[16].

Two recent studies[12,18] have reported zero-resistance in the $La_3Ni_2O_7$. However, it is clearly stated in Ref.[18] that not all highly compressed $La_3Ni_2O_{7-\delta}$ samples exhibit zero-resistance transition. In addition, Zhou *et al*[18] reported the temperature dependent AC susceptibility data, from which it was estimated the presence of the superconducting phase at the volume level of 1% in the sample compressed at $P \geq 20\ GPa$. This result demonstrates that there is a quest to find an intriguing unknown parameter, which determines the appearance of the zero-resistance phase in highly compressed $La_3Ni_2O_{7-\delta}$.

Here, we contributed to the exploration and focused on a detailed analysis of available experimental data measured in highly compressed $La_3Ni_2O_{7-\delta}$ single crystals. While the majority of theoretical groups (but not all[19,20]) explore hypotheses for unconventional mechanisms of pairing in the $La_3Ni_2O_7$, we investigated a possibility for the electron-phonon pairing mechanism.

To do this, we extracted:
(1) pressure dependent Debye temperature, $\Theta_D(P)$;
and based on that, we determined:

(2) the electron-phonon coupling constant, $\lambda_{e-ph}$, for one sample exhibited the zero-resistance state.

In addition, we estimated:

(3) the crystal lattice strain, $\varepsilon(P)$, in the La$_3$Ni$_2$O$_{7-\delta}$ at nanoscale level.

Primary idea to determine the lattice strain, $\varepsilon$, was initiated by recent FPC result by Sanna *et al*[21] who reported that the record-high $T_c$ in titanium[22,23] can be explained within electron-phonon phenomenology, if an assumption about the presence of the vacancies in the crystal lattice can be made. In addition, Liu *et al*[24] performed the FPC studies and concluded that the presence of the apical-oxygen vacancies should dramatically suppress superconducting transition temperature in La$_3$Ni$_2$O$_{7-\delta}$. On the other hand, vacancies are well-known structural imperfections which impact the superconducting state in cuprates[25,26] and iron-based superconductors[27]. It should be also noted that there is some experimental evidence that the lattice strain influences the superconducting transition in doped $R$NiO$_2$ ($R$ = La, Pr, Nd) films[28].

The clearest expressions that the FPC studies should change its primary research object from the ideal defect-free crystal lattices to lattices with defects and strain has been reported by Cucciari *et al*[29].

While there are no experimental techniques which can be used for direct observation of the vacancies or other defects for samples in DAC, here we utilized the Williamson-Hall (WH) analysis [30] of the XRD data to extract the lattice strain, $\varepsilon$, in La$_3$Ni$_2$O$_{7-\delta}$. Advanced WH analysis [31–33] can be used to extract several microstructural parameters. However, here we used classical WH approach [30] to extract the lattice strain, $\varepsilon$, only, because of the high anisotropic crystallographic nature of the La$_3$Ni$_2$O$_7$ lattice, unknown Burgers vectors, $b$, and other unknown structural parameters are required for the advanced analysis.

It should be mentioned that Ren *et al* [34] showed that the lattice strain impacts the $T_{c,onset}$ in doped RNiO$_2$ thin films. In addition, Cui *et al*[35] reported that tensile strain stabilizes high-order Ruddlesden–Popper (RP) nickelates (La$_{n+1}$Ni$_n$O$_{3n+1}$) nickelates, whereas compressive strain favours to the low-order La$_3$Ni$_2$O$_7$ (which is La$_{n+1}$Ni$_n$O$_{3n+1}$ at n = 2). More recently, Oppliger *et al*[36] discovered a new phase of the infinite layer nickelate exhibited giant unit cell. It should be also noted, that microcrystalline strain/phase transitions in superconductors can be induced not only by high-pressure, but also by ionizing irradiation[37,38].

All of above facts are confirmations that the evolution of the $\varepsilon(P)$ is important parameter which needs to be determine in any high-pressure superconductor.

## 2. Experimental data sources

In this study, we analysed experimental datasets reported by Sun *et al*[14]. We also analysed the $R(T)$ curve reported by the same research group in Fig. S9[39] and determined the $\lambda_{e-ph}(P)$ for the sample with a zero-resistance state. Utilized models and mathematical routine for the analysis described within each section. All fits performed by our own codes created in the Origin software. We used one of the standard data weighting methods, which is the methos where the weight is proportional to the 1/standard deviation ratio (this method named "Instrumental" in the Origin software).

## 3. Results and Discussion

Standard technique to determine the Debye temperature, $\Theta_D$, is the fit of the specific heat measurements to Debye model. However, this technique cannot be used in studies of highly compressed conductors because of negligible sample thermal mass in comparison with the DAC mass. However, the fit of $R(T)$ data using the saturated resistance model[40,41] allows for the deduction of $\Theta_D$ as a free-fitting parameter:

$$R(T) = \cfrac{1}{\cfrac{1}{R_{sat}} + \cfrac{1}{R_0 + A \times \left(\cfrac{T}{\Theta_D}\right)^5 \times \int_0^{\frac{\Theta_D}{T}} \frac{x^5}{(e^x-1)(1-e^{-x})}dx}} \quad (1)$$

where $R_{sat}$, $R_0$, $\Theta_D$ and $A$ are free-fitting parameters.

In Fig. 2, we showed the $R(T,P)$ curves and data fits to Eq. 1 for samples *Run 1* [14]. In Fig. S1 (Supplementary Material), Fig. S2, and Fig. S3, we showed the $R(T,P)$ curves and data fits to Eq. 1 for samples *Run 2,3,4* [14], respectively. R-square coefficient of determination ($COD$) for each fit is given in each figure caption.

In Fig. 3,a we summarized all deduced $\Theta_D(P)$ values for all samples for which $R(T,P)$ datasets reported in Ref.[14].

One can see that in the pressure range where the high-pressure *Fmmm* phase exists, the Debye temperature is more or less constant with the approximate value of $\Theta_D = 550\ K$.

Considering that Sun *et al*[14] reported that $T_{c,onset}(P)$ is practically unchanged for pure *Fmmm* phase, a hypothesis about the electron-phonon mediated superconductivity in La$_3$Ni$_2$O$_{7-\delta}$ remains its validity, until more experimental data will be available.

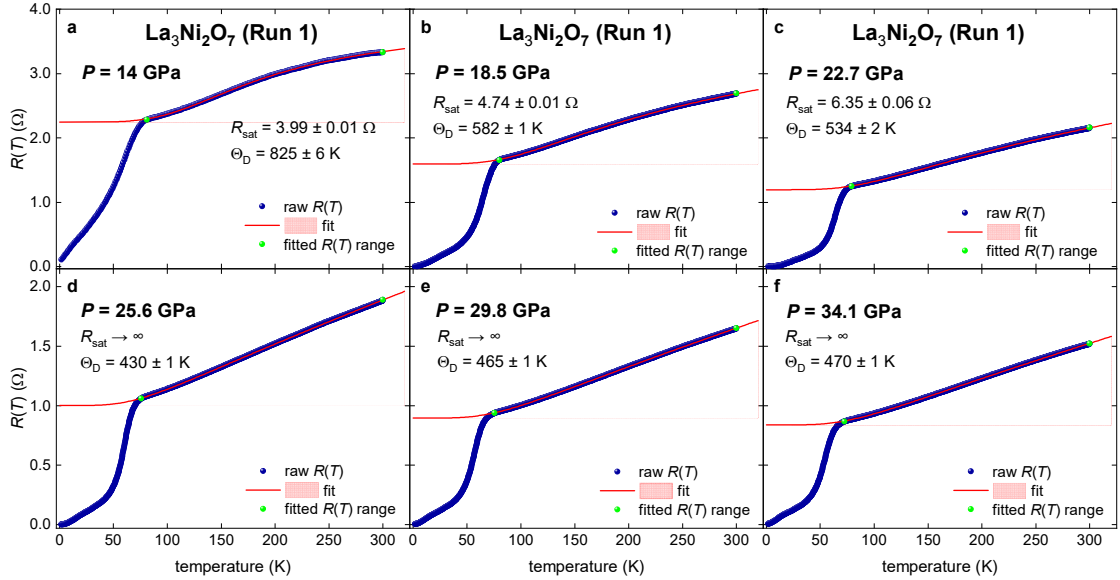

Fig. 2. Temperature dependent resistance, $R(T,P)$, measured in compressed single crystal La$_3$Ni$_2$O$_7$ (Run 1) and data fits to Eq. 1. Raw data reported by Sun *et al*[14]. Green balls indicate the bounds for which $R(T)$ data was used for the fit to Eq. 1. Fit quality for all panels is better or equal to 0.9999. 95% confidence bands are shown by pink areas. (a) $P = 14\ GPa$; (b) $P = 18.5\ GPa$; (c) $P = 22.7\ GPa$; (d) $P = 25.6\ GPa$; (e) $P = 14\ GPa$; (f) $P = 18.5\ GPa$.

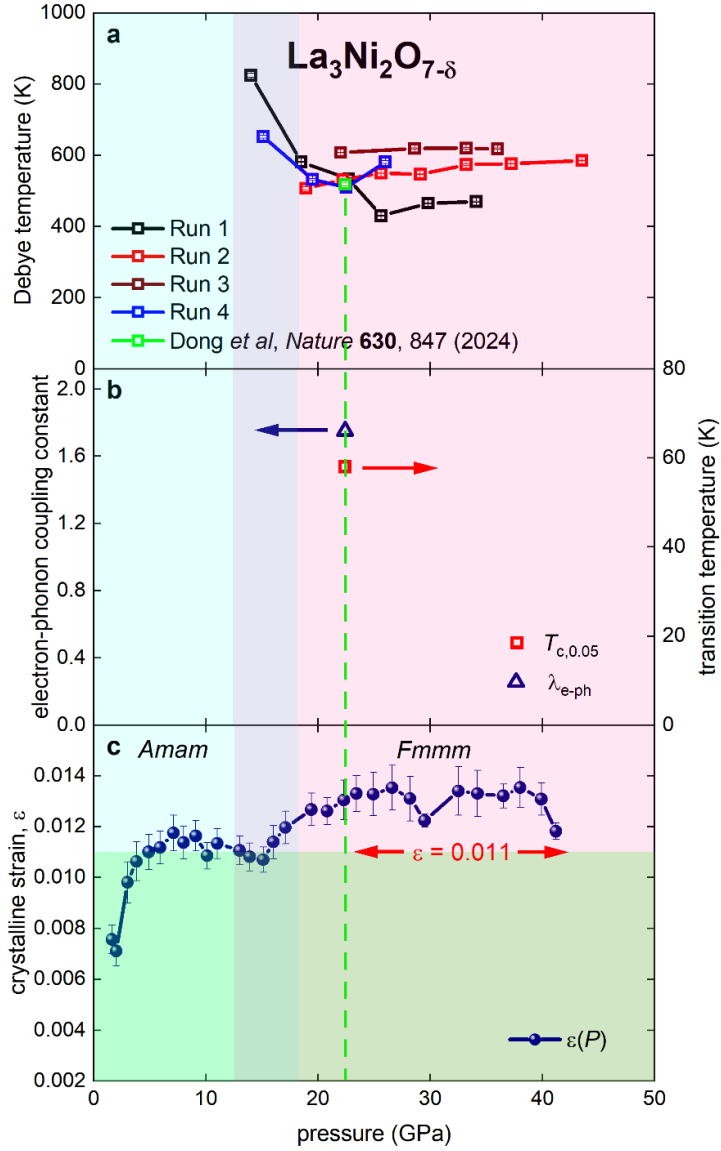

Fig. 3. Evolution of the (a) Debye temperature, $\Theta_D(P)$ on applied pressure; (b) calculated electron-phonon coupling constant, $\lambda_{e-ph}(P = 22.4\ GPa)$, and transition temperature defined by $T_{c,0.05}(P = 22.4\ GPa)$ criterion; and (c) crystalline strain, $\varepsilon(P)$, in single crystal La3Ni2O7-δ. Phase boundaries for the *Amma* and *Fmmm* phases are shown by magenta and cyan areas based on the estimated values reported by Sun *et al*[14].

From deduced $\Theta_D$ and known $T_c$, the electron-phonon coupling constant, $\lambda_{e-ph}$, can be determined as the root of advanced McMillan equation[42]:

$$T_c = \left(\frac{1}{1.45}\right) \times \Theta_D \times e^{-\left(\frac{1.04(1+\lambda_{e-ph})}{\lambda_{e-ph}-\mu^*(1+0.62\lambda_{e-ph})}\right)} \times f_1 \times f_2^*, \quad (2)$$

where

$$f_1 = \left(1 + \left(\frac{\lambda_{e-ph}}{2.46(1+3.8\mu^*)}\right)^{3/2}\right)^{1/3}, \quad (3)$$

$$f_2^* = 1 + (0.0241 - 0.0735 \times \mu^*) \times \lambda_{e-ph}^2, \quad (4)$$

where $\mu^*$ is the Coulomb pseudopotential. In this work we assumed that $\mu^* = 0.13$, which is a typical value for highly compressed electron-phonon mediated superconductors[8].

Considering all issues mentioned in the Introduction regarding the zero-resistance problem in nickelates, here we analysed the $R(T, P = 22.4\ GPa)$ measured in single crystal La$_3$Ni$_2$O$_{7-\delta}$[39], and in which the resistance reduces to undistinguishable, from measurement system noise, level. To extract $\Theta_D$ and $T_c$, we utilized full $R(T)$ curve fitting[43]:

$$R(T) = \frac{1}{\frac{\theta(T_c^{onset}-T)}{\left(I_0\left(F\times\left(1-\frac{T}{T_c^{onset}}\right)^{3/2}\right)\right)^2} + \theta(T-T_c^{onset})\times\left(\frac{1}{R_{sat}}+\frac{1}{\left(R_0+A\times\left(\frac{T}{\Theta_D}\right)^5\times\int_0^{\frac{\Theta_D}{T}}\frac{x^5}{(e^x-1)(1-e^{-x})}dx\right)}\right)}, \quad (5)$$

where $\theta(x)$ is the Heaviside step function, $I_0(x)$ is the zero-order modified Bessel function of the first kind, $R_0(T_c^{onset})$, $T_c^{onset}$, $F$, $R_{sat}$, $R_0$, $\Theta_D$, and $A$ are free-fitting parameters. We defined the transition temperature by the criterion:

$$\left.\frac{R(T)}{R(T_{c,onset})}\right|_{T_{c,0.05}} = 0.05 \quad (6)$$

The fit is shown in Fig. 4. It should be stressed that derived $\Theta_D(22.4\ GPa) = 518 \pm 4\ K$ is practically the same for four of five analysed $R(T)$ datasets showed in Fig. 3.

Derived $T_{c,0.05} = 58.0 \pm 0.1\ K$ and $\lambda_{e-ph} = 1.75$. Deduced $\lambda_{e-ph}$ shows, that if the high-temperature superconducting state in La$_3$Ni$_2$O$_{7-\delta}$ originates from the electron-phonon interaction, this requires the interaction strength at its upper limit, similar to the interaction strength exhibited in highly compressed hydrides[8,44–47].

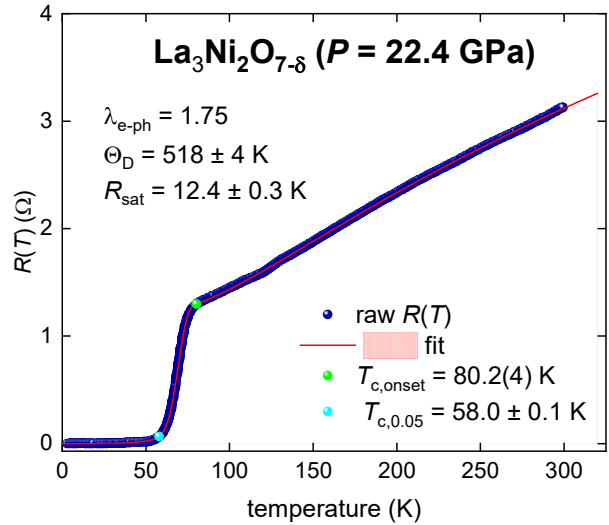

Fig. 4. Temperature dependent resistance, $R(T,P=22.4\ GPa)$, in compressed single crystal La$_3$Ni$_2$O$_{7-\delta}$ and data fit to Eq. 5. Raw data reported by Dong *et al*[39] Fit quality is 0.99994.

Derived $\lambda_{e-ph} = 1.75$ value is close to $\lambda_{e-ph} = 1.70$ of Nb$_3$Al and Nb$_3$Sn[48], and this value is significantly lower than the $\lambda_{e-ph} = 2.40$ reported for ambient pressure oxide superconductor KOs$_2$O$_6$[49].

XRD peaks [14] were approximated by Lorentz function:

$$I(2\theta) = I_{background} + \sum_{k=1}^{N}\frac{2\times I_k}{\pi}\times\frac{\frac{2}{\pi}\beta_k}{4\times(2\theta-2\theta_{peak,k})^2+\left(\frac{2}{\pi}\beta\right)_k^2}, \quad (7)$$

where $I_k$ is the peak area, $2\theta_{peak,k}$ is the peak position, $\beta_k$ is peak integral breadth, and $I_k$, $2\theta_{peak,k}$, and $\beta_k$ are-free fitting parameters. $I_{background}$ level was chosen manually in all fits. Details can be found elsewhere [50].

Derived peaks breadth, $\beta(\theta)$, and peaks diffraction angle, $\theta$, were fitted to classical Williamson-Hall (WH) model[30] (where we assumed that the instrumental broadening, $\beta_i$, is negligible):

$$\beta(\theta, P) = \frac{0.9 \times \lambda_{X-ray}}{D(P) \times cos(\theta)} + 4 \times \varepsilon(P) \times tan(\theta), \quad (8)$$

where $\lambda_{X-ray} = 61.99\ pm$ is the wavelength of used radiation in Ref.[14], and $D(P)$ is the mean size of coherent scattering regions at a given pressure $P$. As we mentioned above, in these fits, we used the weighting method of 1/standard deviation.

Performed fits showed that for all pressures, $1.6\ GPa \leq P \leq 41.2\ GPa$, the size of coherent scattering regions, $D(P)$, is large and the uncertainty of the value exceeds the value itself by far. Thus, we fit data to the reduced equation:

$$\beta(\theta, P) = 4 \times \varepsilon(P) \times tan(\theta). \quad (8)$$

We show some fits in Fig. S4, and we summarised results in Figs. 3,c, and 5, where one can see that the $\varepsilon(P)$ is raising reasonably steep at low applied pressure, up to $P = 4.9\ GPa$.

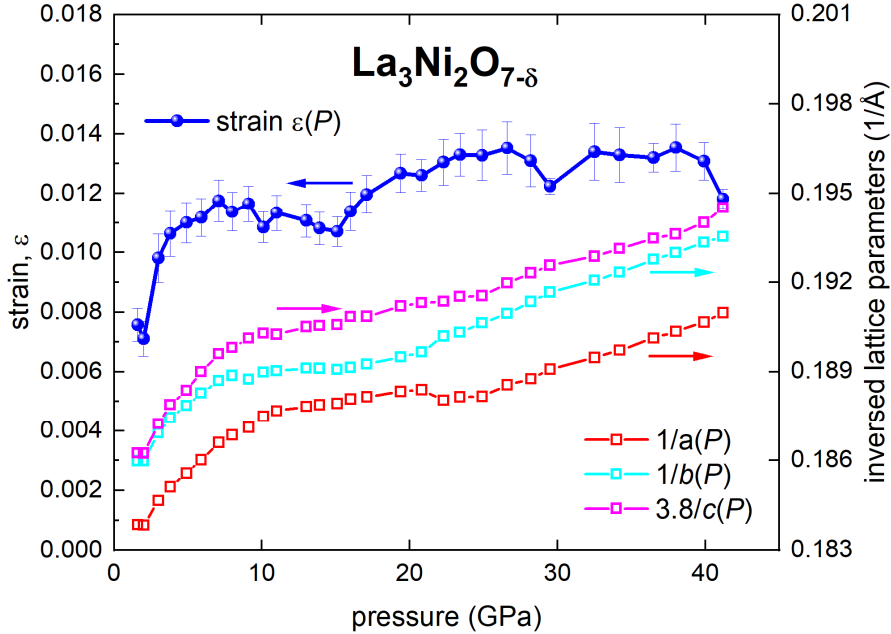

Fig. 5. Crystalline strain, $\varepsilon(P)$, and inverse lattice constants $a(P)$, $b(P)$, and $c(P)$ dependence from applied pressure in highly compressed single crystal La$_3$Ni$_2$O$_{7-\delta}$. Raw data for $a(P)$, $b(P)$, and $c(P)$ reported by Sun *et al*[32].

We detected two deeps in the $\varepsilon(P)$ at $P = 15.1\ GPa$ and $29.5\ GPa$ (Figs. 3,c and 5). Our analysis showed that ε(P) has three-dome shape in the pressure range of 1.6 GPa ≤ P ≤ 41.2 GPa with peaks located at $P \cong 8, 25, and\ 36\ GPa$ (Figs. 3,c and 5).

The first deep at $P = 15.1\ GPa$ coincides with the transition pressure from the *Amam* into the *Fmmm* phase, and perhaps it is direct structural evidence of the lattice relaxation at the structural transition.

However, the $\varepsilon(P)$ dependence does not exactly match the lattice constants dependences (see, $1/a(P)$, $1/b(P)$, and $1/c(P)$ in Fig. 5), especially at low- and high-$P$ ranges. This difference, in particular, at low-$P$ looks illogical.

However, we can explain the latter because of reducing the lattice volume by reducing the volume of each vacancy. This reduction can occur without significant changes in the lattice strain, because the vacancies density remains the same.

Perhaps, high vacancies density can be the origin for the observation/absence of the zero-resistance state in the $La_3Ni_2O_{7-\delta}$[14]. Simple fact that there are no sharp simultaneous changes in $a(P)$, $b(P)$, $c(P)$, and $\varepsilon(P)$, except, perhaps, the change in curves slope at $P \sim 20\ GPa$ is an indication that the phase transition *Amam-Fmmm* is very wide.

We also need to stress that all XRD datasets (which we analysed) were collected at room temperature[14]. Obviously, that at the temperature range from $T = 300\ K$ down to $T = 50\ K$ some phase structural transition, or multiple transitions, can occur.

Another important revealed issue is that the level of the microcrystalline strain, $\varepsilon(P)$, in the *Amam*-phase of the $La_3Ni_2O_{7-\delta}$ samples is higher than the one measured in perfect undoped $YBa_2Cu_3O_{7-\delta}$ films[51,52], $0.0039 \leq \varepsilon_{YBCO\ film} \leq 0.0078$. Thus, the *Fmmm*-phase exhibits significantly higher microcrystalline strain in comparison with undoped $YBa_2Cu_3O_{7-\delta}$. It would be interesting to determine the microcrystalline strain in recently discovered iron-based [53-56] and hydrogen-based superconductors [57-60] to build generic picture on the impact of microcrystalline stain on superconducting properties of high-temperature superconductors. First study of the crystalline strain at nanoscale level in hydrogen-based superconductor $La_4H_{23}$ (with $T_{c,zero} \sim 70$ K) has performed recently [50]. This study [50] showed that the $La_4H_{23}$ has a low level of strain $|\varepsilon(P)| \leq 0.003$.

Based on all above, we should note that there is a need for low-temperature high-pressure XRD studies of $La_3Ni_2O_{7-\delta}$, which can be used for detailed analysis of the phase transition(s) and related structural/phase parameters.

### 4. Conclusions

Highly compressed $La_3Ni_2O_{7-\delta}$ is under intensive experimental and theoretical studies at the moment [11-14,20,35,36,39,61,62]. In this work, we analyzed experimental data and determined:

(1) pressure dependent Debye temperature, $\Theta_D(P)$;

(2) three-dome shape of the $\varepsilon(P)$ in the studied pressure range, where one deep of the $\varepsilon(P)$ coincides with the pressure at which the *Amam* into the *Fmmm* phase transition occurs; and

(3) electron-phonon coupling constant, $\lambda_{e-ph}(P = 22.4\ GPa) = 1.75$;

in single crystal $La_3Ni_2O_{7-\delta}$.

*Supplementary Material*

In Supplementary Materials we showed analysis for temperature dependent resistance data, and XRD data.

*Acknowledgements*

The authors thank financial support provided by the Ministry of Science and Higher Education of Russia (theme "Pressure" No. 122021000032-5, and theme "Spin" No. 122021000036-3).

*Supplementary Materials*

# Debye temperature, electron-phonon coupling constant, and three-dome shape of crystalline strain as a function of pressure in highly compressed La$_3$Ni$_2$O$_{7-\delta}$

Evgeny F. Talantsev[†][1,2] and Vasiliy V. Chistyakov[1,2]

[†]e-mail: e.f.talantsev@gmail.com

[1]M.N. Miheev Institute of Metal Physics, Ural Branch, Russian Academy of Sciences, 18, S. Kovalevskoy St., Ekaterinburg, 620108, Russia

[2]NANOTECH Centre, Ural Federal University, 19 Mira St., Ekaterinburg, 620002, Russia

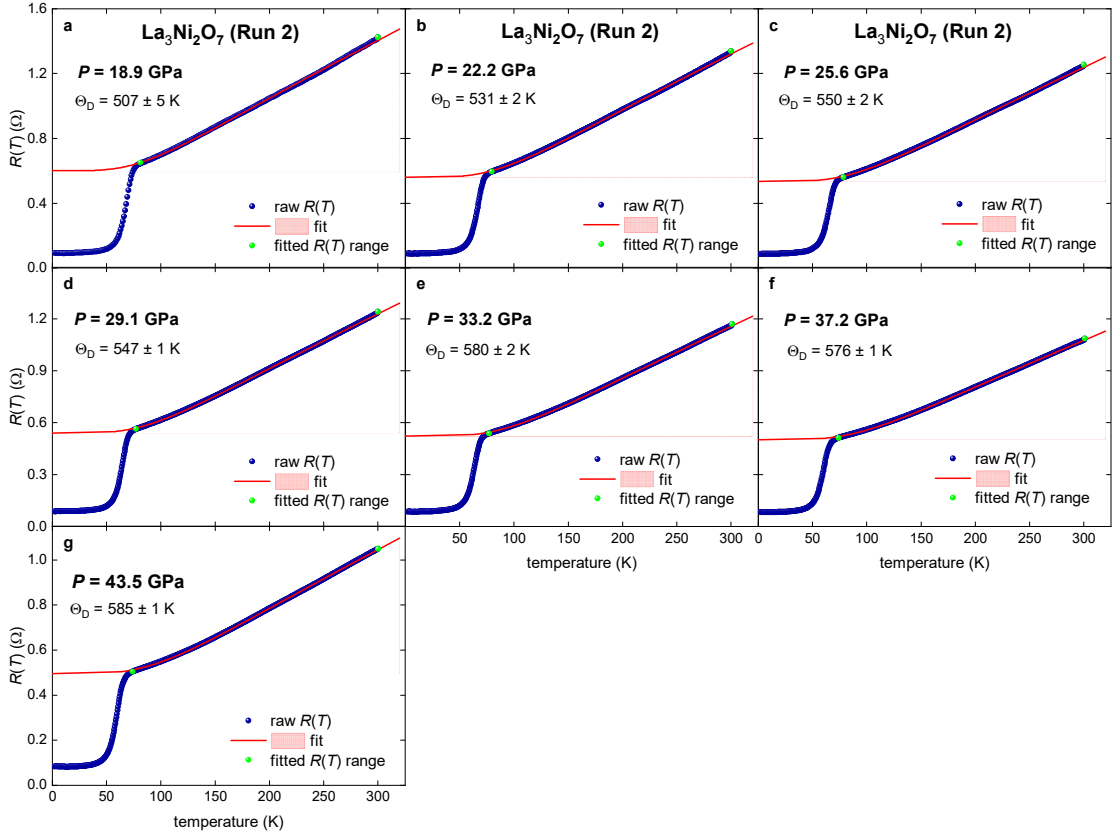

**Figure S1.** Temperature dependent resistance, $R(T,P)$, measured in compressed single crystal La$_3$Ni$_2$O$_7$ (*Run 2*) and data fits to Eq. 1. Raw data reported by Sun *et al*[32]. $R_{sat} \to \infty$ for all fits. Green balls indicate the bounds for which $R(T)$ data was used for the fit to Eq. 1. Fit quality for all panels is better or equal to 0.9999. 95% confidence bands are shown by pink areas. (a) $P = 18.9\ GPa$; (b) $P = 22.2\ GPa$; (c) $P = 25.6\ GPa$; (d) $P = 29.1\ GPa$; (e) $P = 33.2\ GPa$; (f) $P = 37.2\ GPa$; (g) $P = 43.5\ GPa$.

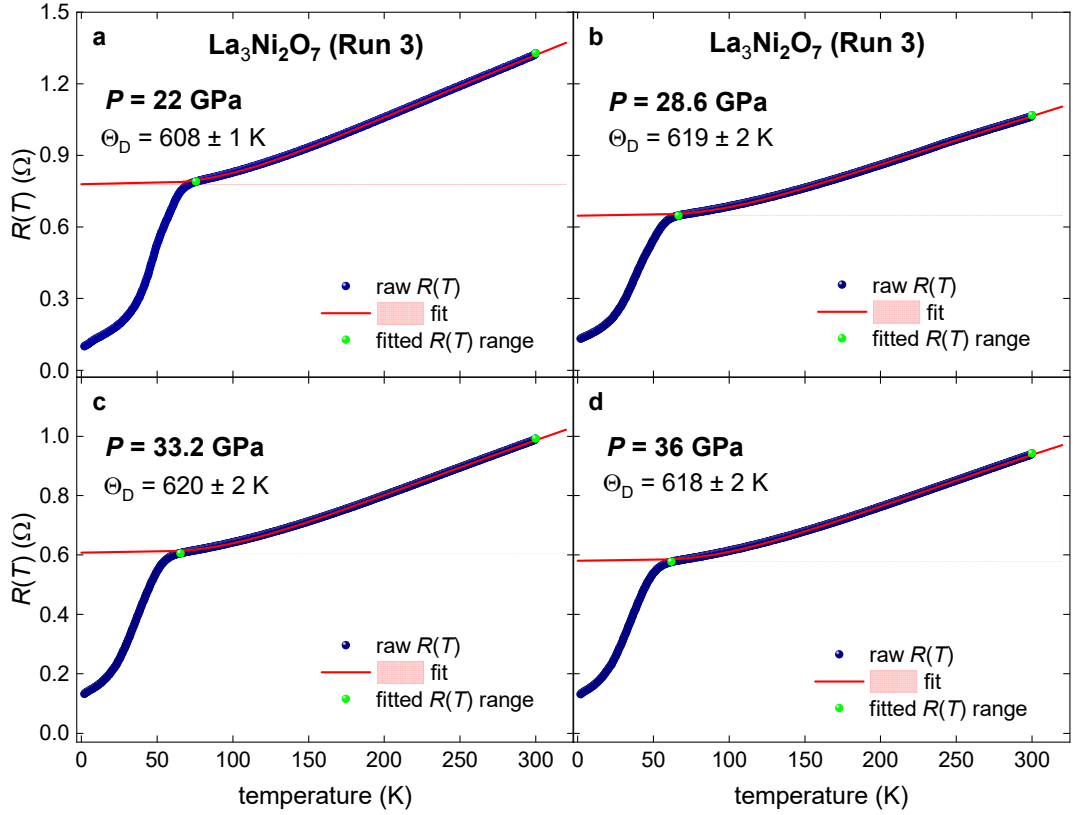

**Figure S2.** Temperature dependent resistance, $R(T,P)$, measured in compressed single crystal $La_3Ni_2O_7$ (*Run 3*) and data fits to Eq. 1. Raw data reported by Sun *et al*[32]. $R_{sat} \to \infty$ for all fits. Green balls indicate the bounds for which $R(T)$ data was used for the fit to Eq. 1. Fit quality for all panels is better or equal to 0.9997. 95% confidence bands are shown by pink areas. (a) $P = 22\ GPa$; (b) $P = 28.6\ GPa$; (c) $P = 33.2\ GPa$; (d) $P = 36\ GPa$.

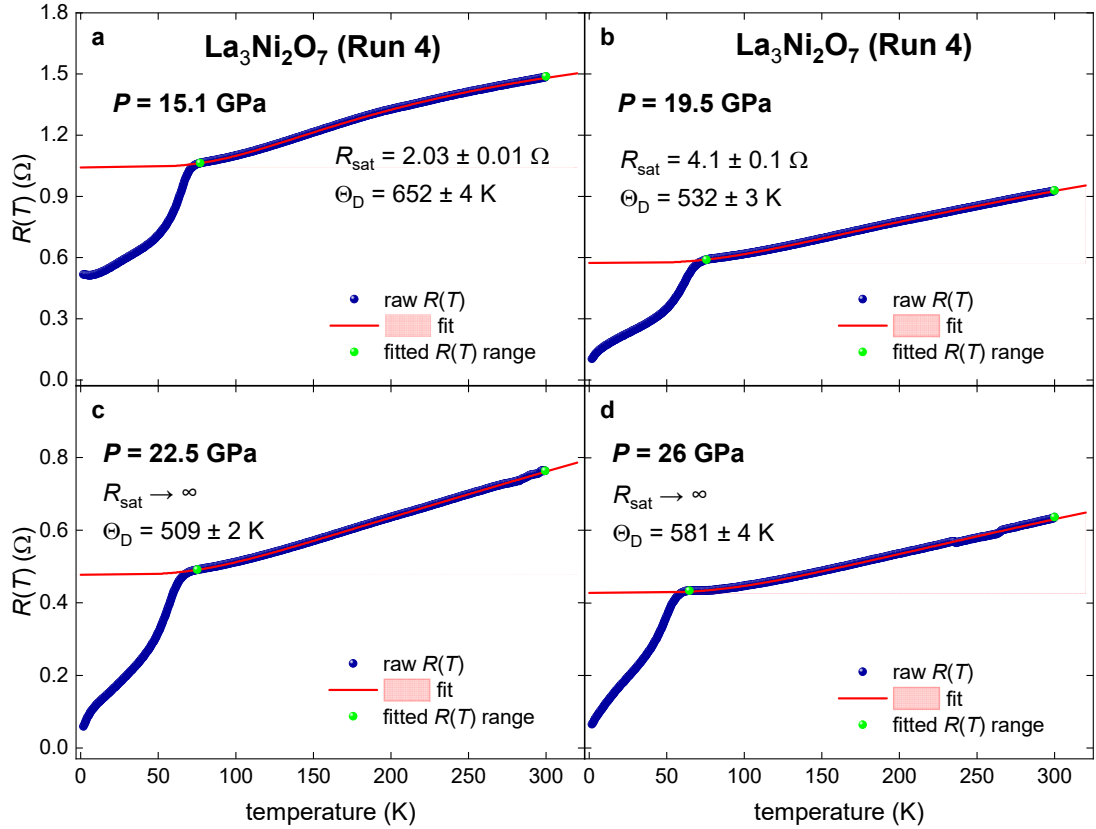

**Figure S3.** Temperature dependent resistance, $R(T,P)$, measured in compressed single crystal La$_3$Ni$_2$O$_7$ (*Run 4*) and data fits to Eq. 1. Raw data reported by Sun *et al*[32]. Green balls indicate the bounds for which $R(T)$ data was used for the fit to Eq. 1. Fit quality for all panels is better or equal to 0.9989. 95% confidence bands are shown by pink areas. (a) $P = 15.1\ GPa$; (b) $P = 19.5\ GPa$; (c) $P = 22.5\ GPa$; (d) $P = 26\ GPa$.

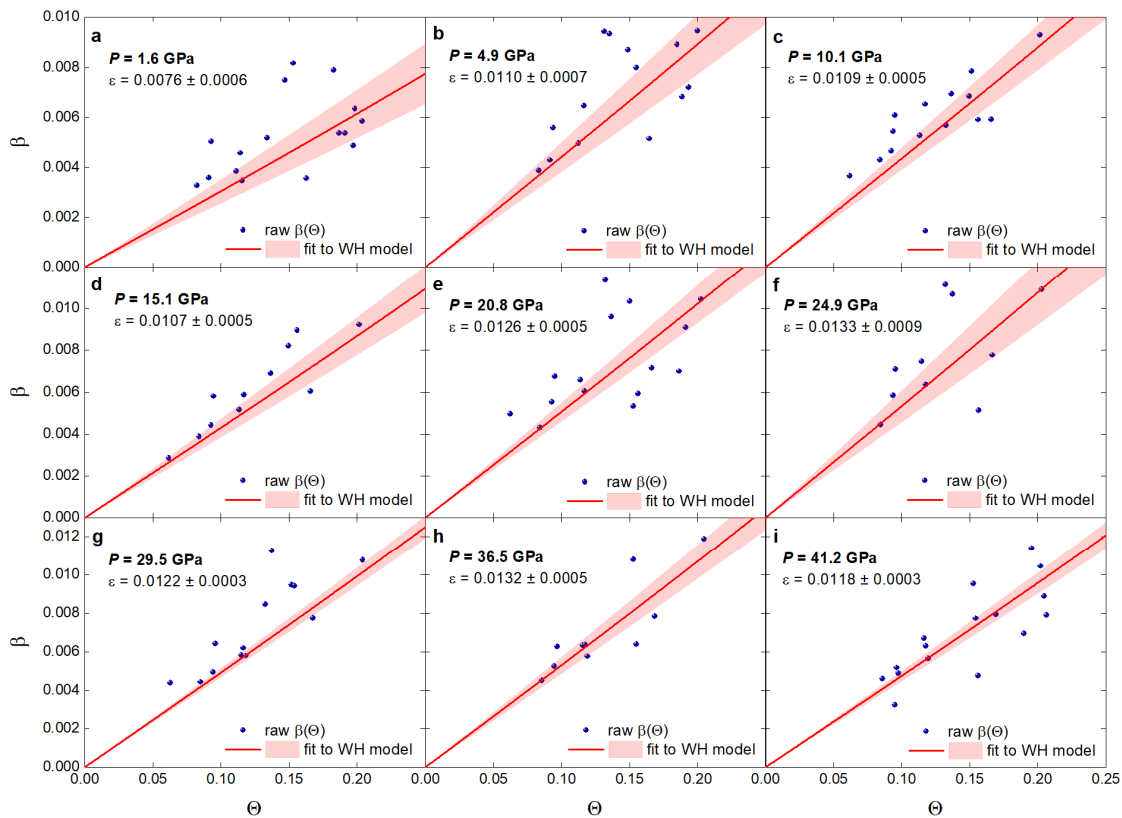

**Figure S4.** XRD peaks breadth, $\beta(\theta)$, fits to reduced Williamson-Hall model (Eq. 8) for highly compressed single crystal $La_3Ni_2O_{7-\delta}$. Raw XRD scans reported by Sun *et al*[32]. 95% confidence bands are shown by pink areas. (a) $P = 1.6\ GPa$; (b) $P = 4.9\ GPa$; (c) $P = 10.1\ GPa$; (d) $P = 15.1\ GPa$, (e) $P = 20.8\ GPa$; (f) $P = 24.9\ GPa$; (g) $P = 29.5\ GPa$; (h) $P = 36.5\ GPa$, (i) $P = 41.2\ GPa$.